# Mass-like gap creation by mixed singlet and triplet state in superconducting topological insulator

M. Khezerlou[*,1,2], and H. Goudarzi[†1]

[1]*Department of Physics, Faculty of Science, Urmia University, P.O.Box: 165, Urmia, Iran*

[2]*National Elites Foundation, Iran*

**Abstract**

We investigate proximity-induced mixed spin-singlet and spin-triplet superconducting state on the surface states of a topological insulator. Such hybrid structure features fundamentally distinct electron-hole excitations and resulting effective superconducting subgap. Studying the particle-hole and time-reversal symmetry properties of the mixed state Dirac-Bogoliubov-de Gennes effective Hamiltonian gives rise to manifesting possible topological phase exchange of surface states, since the mixed-spin channels leads to appearance of a band gap on the surface states. This is verified by determining topological invariant winding number for chiral eigenstates, which is achieved by introducing a chiral symmetry operator. We interestingly find the role of mixed superconducting state as creating a mass-like gap in topological insulator by means of introducing new mixed-spin channels $\Delta_1$ and $\Delta_2$. The interplay between superconducting spin-singlet and triplet correlations actually results in gaped surface states, where the size of gap can be controlled by tuning the relative $s$ and $p$-waves pairing potentials. We show that the system is in different topology classes by means of chiral and no-chiral spin-triplet symmetry. In addition, the resulting effective superconductor subgap manipulated at the Fermi surface presents a complicated dependency on mixed-spin channels. Furthermore, we investigate the resulting subgap tunneling conductance in N/S and Josephson current in S/I/S junctions to unveil the influence of effective symmetry of mixed superconducting gap. The results can pave the way to realize the effective superconducting gap in noncentrosymmetric superconductors with mixed-spin state.

## I  INTRODUCTION

Topological insulators (TIs), as an interesting topologically nontrivial phase of condensed matter represent distinct electronic properties comparing to the conventional band insulators. On the surface of a three-dimensional topological insulator (3DTI), topologically protected quantum channels are formed in a manner that the charge carriers obey from massless Dirac-like fermions. These gapless surface states are protected by time-reversal (TR) symmetry and are robust against disorder and perturbations. There exist odd number of Dirac cones in the Brillouin zone, resulted from inversion symmetry breaking owing to the Rashba-type spin-orbit interaction [1, 2]. These peculiar features enable TIs to be potentially used to spintronics [3, 4, 5] and topological quantum information applications [6, 7, 8]. Moreover, superconductivity induction by proximity-effect on the surface states of a 3DTI has been of noticeable importance during the last decade. Several experimental probes [9, 10, 11, 12, 13, 14, 15] have evidenced existence of spin-singlet and spin-triplet pairing states in the hybrid structure of a 3DTI and a superconductor. One of key findings in this topic is the manipulation of Majorana fermions in the Andreev bound states (ABS) established at the 3DTI ferromagnet-superconductor (FS) interface [16, 17, 18]. Merging the spin-singlet

[*]m.khezerlou@urmia.ac.ir

[†]h.goudarzi@urmia.ac.ir

Bogoliubov-de Gennes Hamiltonian with gapless TR symmetric surface states gives rise to appearance of triplet-like components of superconductor gap in the resulting Dirac-Bogoliubov-de Gennes (DBdG) Hamiltonian, originated from the requirement of states to be invariant under particle-hole (PH) symmetry. Consequently, the quasiparticle's energy excitation remains gapless, when the proximity-induced superconducting order parameter is taken to be spin-triplet $p$-wave symmetry. Actually, this leads to suppression of Andreev process for energy excitations lower than superconducting effective gap [19]. Unconventional superconductivity in 2D Dirac materials plays an important role [19, 20, 21, 22].

However, regarding the inversion symmetry breaking in TIs, it will not be (at least from the dynamical symmetry point of view) ungraceful to take into mixed spin-singlet and spin-triplet superconducting state contribution to the quasiparticle excitations, since the $(s + p)$-wave state is also found to break the inversion symmetry. The symmetry of Cooper pair states in new systems with broken inversion symmetry, such as noncentrosymmetric (NCS) superconductors can not be classified based on orbital and spin parts. The Cooper pair in these systems is, therefore, a mixture of singlet and triplet spin states. A new type of NCS superconductor $Zr_3Ir$ has been very recently reported [23]. It is noted that, in cuprats, which have no inversion center in their crystal structure, the inversion symmetry is broken, leading to appearance of robust asymmetric spin-orbit interaction. Therefore, the superconducting pair potential mixes singlet and triplet states [24, 25]. As a noticeable result, the spin-polarized current has been predicted to appear on the surface of a superconductor with mixed singlet and chiral triplet state [26, 27]. In addition, in these materials, there is an even number of Majorana fermions in two-dimensional (2D) TR symmetric superconducting bound states [28, 29].

Moreover, the exact pairing potentials describing many of superconductors with mixed singlet and chiral triplet states remain unknown. An updated list of recently discovered such superconductors can be found in Ref. [30]. As mentioned, 2D materials with spin-orbit coupling, such as 3DTI, is expected to host a triplet state using a conventional $s$-wave superconductor, for example see, Ref. [31]. Interestingly, the surface state of superconducting 3DTI is strongly related to the 2D unconventional superconductors (such as $p$-wave symmetry) with broken inversion symmetry [28]. Hence, the outcomes of transport of charge carriers on the surface of a 3DTI hybrid (with mixed singlet and triplet superconducting states) contact can pave the way to unveil the effective dynamics of superconductor pairing state.

Therefore, we proceed, in this paper, to investigate particularly a newly appeared distinct manifestation of presence of mixed $s$ and $p$-wave symmetries on the surface states (see Fig. 1(a)). From this point of view, the interplay between inversion, PH and TR symmetries in the 3DTI mixed superconductor Hamiltonian is expected to introduce a distinct scenario for quasiparticle excitation, where we find it to present a mass-like gap opening in Dirac point by the varying comparative magnitudes of singlet $s$-wave and triplet $p$-wave pairing potentials. We try to show the possible topological property of band structure via calculating the winding number, which is related to the Berry phase reflecting the topological structure of wavefunction [32]. Specifically, electro-hole conversion at the normal metal-superconductor (NS) interface, called Andreev reflection (AR), and appearance of chiral Majorana state at the ferromagnet-superconductor (FS) interface can be considered as essential phenomena, which is directly influenced by the interplay between the spin-singlet and spin-triplet states on top of a 3DTI [33, 34, 35]. In light of the above attributes, further treatment to study mixed-state 3DTI can be more impressive to evaluate its dynamical and transport properties. We show that the effective superconductor gap at the interface has a more complicated dependency on the magnitude of $s$- and $p$-wave pair potentials ($\Delta_s$ and $\Delta_p$). The Dirac-point gapless band structure with $s$-wave symmetry is converted to the gaped states with the mixed $(s + p)$-wave symmetry via $\mathcal{Z}_2$ topological invariant. In Ref. [36], the authors have just pointed out the energy eigenvalue of hybrid 3DTI mixed superconductor Hamiltonian. Here, we have succeeded to present analytical expression of the corresponding electron(hole) wavefunction in order to capture its topological nature, the resulting Andreev subgap tunneling conductance (Fig. 1(b)) and, of course, the $(0 - \pi)$-current-phase relation in a Josephson junction (Fig. 1(c)). These results have been achieved in a really cumbersome analytical procedure.

This paper is organized as follows. Section II is devoted to describe the discrete symmetry properties of 3DTI Hamiltonian in the presence of mixed superconducting state. The chiral symmetry of the system is investigated. Next, the effective superconducting gap and electron-hole energy excitations are introduced. The winding number for electron-hole pairing in a closed Berry connection in Brillouin zone is studied by using the analytically obtained chiral eigenstates. In Sec. IIIA, we represent the explicit expressions of normal and Andreev reflection amplitudes in a corresponding NS junction. The numerical results of subgap tunneling conductance are presented along with a discussion of main characteristics of system. The Josephson junction is considered in Sec. IIIB in order to investigate the property of Andreev bound state (ABS) and resulting supercurrent-phase exhibition. Finally, a brief discussion is given in Sec. IV.

## II THEORETICAL FORMALISM

### A Discrete symmetries of mixed state 3DTI

We begin by setting up a topological insulator-based model for the proximity effect, that the pairing potential contains both spin-singlet and spin-triplet states. The order parameter for a mixture of such state adopts the general form $\hat{\Delta}_m(\mathbf{k}) = ie^{i\varphi}\left[\Delta_s(\mathbf{k})\hat{\sigma}_0 + \mathbf{d}(\mathbf{k})\cdot\hat{\boldsymbol{\sigma}}\right]\hat{\sigma}_2$, where the Pauli matrices $\hat{\sigma}_i$ acting on the spin space and $\varphi$ indicates the superconducting phase. The spin-singlet component is an even function of the wave vector, and we assume that the pairing potential $\Delta_s(\mathbf{k}) = |\Delta_s|$ to be constant and real. The order parameter of spin-triplet pairing is described by an odd vector function $\mathbf{d}(\mathbf{k})$ of the momentum. For the chiral spin-triplet pairing, $\mathbf{d}(\mathbf{k})$ may then be written in the form $\mathbf{d}(\mathbf{k}) = |\Delta_p|\left[\cos\theta + i\chi\sin\theta\right]\hat{z}$, where $|\Delta_p|$ measures the amplitude of the triplet order parameter and $\chi$ labels the orientation of the angular momentum of the Cooper pair (featuring the chirality).

The real and positive parameter $\Delta_m$ is introduced to quantify the energy scale of the superconducting gap. Throughout present work, the singlet $\Delta_s$ and triplrt $\Delta_p$ pair potential parameters are normalized by $\Delta_m$. We employ the DBdG Hamiltonian

$$\mathcal{H}(\mathbf{k}) = \begin{pmatrix} \hat{h}_{TI}(\mathbf{k}) & \hat{\Delta}_m(\mathbf{k}) \\ \hat{\Xi}\hat{\Delta}_m(\mathbf{k})\hat{\Xi}^{-1} & \hat{\Xi}\hat{h}_{TI}(\mathbf{k})\hat{\Xi}^{-1} \end{pmatrix} \quad (1)$$

in Nambu space for the surface states of a topological insulator to obtain the energy dispersion relation under the influence of superconducting proximity effect. The gapless surface states are described by the 2D linear Hamiltonian $\hat{h}_{TI}(\mathbf{k}) = v_F\left(\hat{\sigma}_1 k_x + \hat{\sigma}_2 k_y\right) - \mu_s$, ($\hbar = 1$), where $v_F$ and $\mu_s$ denote velocity of charge carriers and chemical potential, respectively. PH symmetry operator $\hat{\Xi}$ is involved by an antiunitary operator, which may act on Dirac Hamiltonian and superconductor order parameter. By acting the PH symmetry operator and defining two complex pair potentials as $\Delta_{1,2} = \Delta_s \pm \Delta_p\, e^{i\chi\theta}$, the $4\times 4$ Hamiltonian of mixed superconducting topological insulator hybrid yields:

$$\mathcal{H}(\mathbf{k}) = \begin{pmatrix} -\mu_s & v_F|k|\,e^{-i\theta} & 0 & \Delta_1 \\ v_F|k|\,e^{i\theta} & -\mu_s & -\Delta_2 & 0 \\ 0 & -\Delta_2^* & \mu_s & v_F|k|\,e^{i\theta} \\ \Delta_1^* & 0 & v_F|k|\,e^{-i\theta} & \mu_s \end{pmatrix}. \quad (2)$$

Spin-singlet and spin-triplet admixture gives rise to two new spin channels $\Delta_1$ and $\Delta_2$. The effective mixed pairing potential depends on the angle $\theta$, where, only for $\Delta_2$ channel, there exist the possibility to be zero. This case occurs when spin-triplet contribution is dominated, $|\Delta_p| \geq |\Delta_s|$. Both spin channels $\Delta_1$ and $\Delta_2$ have no zero value for every angle $-\pi/2 \leq \theta \leq \pi/2$, when spin-singlet potential is dominant. The effective two mixed-spin pair potentials is demonstrated, in detail, in Fig. 1(d).

Let now unveil the topological symmetry properties of this given state. The resulting effective Hamiltonian (2) satisfies the PH symmetry relation, which is $-\mathcal{H}^*(-\mathbf{k}) = \hat{\Xi}\mathcal{H}(\mathbf{k})\hat{\Xi}^{-1}$, when

$$\hat{\Xi} = (\hat{\tau}_1 \otimes \hat{\sigma}_0)\hat{\mathcal{C}}.$$

$\hat{\mathcal{C}}$ is the complex conjugation operator. The operator $\hat{\tau}_1$ is the Pauli matrix in particle-hole space. In this case, the needed PH symmetry of mixed superconductor gap is provided in the surface states of 3DTI. The square PH symmetry operator is found $\hat{\Xi}^2 = +1$. It is noted, that this symmetry may prove the spin degeneracy of the Fermi surface to be lifted, and consequently it allows for exotic chiral Majorana modes [37]. On the other hand, the TR symmetry operator can be given by

$$\hat{\Theta} = (\hat{\tau}_0 \otimes i\hat{\sigma}_2)\hat{\mathcal{C}}$$

(with $\hat{\tau}_0$ being in particle-hole space), under which the Hamiltonian $\mathcal{H}(\mathbf{k})$ is related to $\mathcal{H}^*(-\mathbf{k})$. Note that, the presence of chiral spin-triplet pairing causes TR symmetry breaking.

By means of specific topological invariant, we remember that, in each spatial dimension, there exist five distinct classes of topological insulators, three of which are characterized by an integral $\mathcal{Z}$ topological number, while the remaining two possess a binary $\mathcal{Z}_2$ topological quantity. Regarding the particle-hole and chirality symmetries of matrices associated with the proposed Hamiltonian, one can determine the topology class. According given topological classification in Ref. [32], the Hamiltonian (2) is found to be placed in topologically nontrivial symmetry class *D*. Meanwhile, for two other no chiral case of spin-triplet *p*-wave symmetry ($\mathbf{d}(\mathbf{k}) = \Delta_p \cos\theta \hat{z}$ and $\Delta_p \sin\theta \hat{z}$), we find Hamiltonian to commute with TR symmetry operator. Hence, the new topology class can be possible, and the system is classified in topology class *DIII* [32].

## B  Mass-like gap

The energy dispersion relation for superconducting excitations can be obtained by diagonalizing the Eq. (2). It is instructive to diagonalize the Hamiltonian $\mathcal{H}$ upon a unitary transformation $\mathcal{H}' = \hat{U}\mathcal{H}\hat{U}^\dagger$. We introduce a unitary matrix to do this goal

$$\hat{U} = \frac{1}{\sqrt{2}} \begin{pmatrix} \hat{\sigma}_0 & \hat{\sigma}_1 \\ \hat{\sigma}_1 & -\hat{\sigma}_0 \end{pmatrix}, \tag{3}$$

under which $\mathcal{H}'$ is transformed to block-diagonal form

$$\mathcal{H}'(\mathbf{k}) = \begin{pmatrix} \Delta_1 & v_F|k|\,e^{-i\theta} & 0 & 0 \\ v_F|k|\,e^{i\theta} & -\Delta_2 & 0 & 0 \\ 0 & 0 & \Delta_2 & v_F|k|\,e^{i\theta} \\ 0 & 0 & v_F|k|\,e^{-i\theta} & -\Delta_1 \end{pmatrix}.$$

The presence of mixed two spin channels $\Delta_1$ and $\Delta_2$ in diagonal elements implies appearance of band gap on the surface states of 3DTI. Hence, the energy eigenvalue dependency on the mixed spin channels is easily given by

$$\mathcal{E}_{mix}(\Delta_1, \Delta_2) = \zeta \sqrt{(v_F|k|)^2 + \tilde{\Delta}_1 + \mu_s^2 + \upsilon\sqrt{\varepsilon_R}}, \tag{4}$$

where $\varepsilon_R = \tilde{\Delta}_2^2 + (2v_F|k|\mu_s)^2 + (v_F|k|)^2 |\Delta_1 - \Delta_2|^2$ denotes the renormalized excitation energy related to the mixed state. $\zeta = \pm 1$ refers the electron-like and hole-like excitations, and $\upsilon = \pm 1$ distinguishes between the conduction and valence bands. The parameters $\tilde{\Delta}_1$ and $\tilde{\Delta}_2$ are defined $\tilde{\Delta}_{1,2} = \frac{1}{2}\left(|\Delta_1|^2 \pm |\Delta_2|^2\right)$ as new normalized mixed spin channels.

Simple inspection of the electron-hole excitation spectrum in NCS superconductors indicates, that there is an essential physical distinction in surface states of topological insulators with mixed pairing state. The Hamiltonian of two-dimensional NCS superconductors is decoupled into two spin channels $\Delta_1$ and $\Delta_2$ with different energies. The exchange between two energies is provided by the sign of electron wave vector [25]. Hence, the pairing potential is only related to the direction of motion (i.e. $\pm|k|$). In the presence of a topological insulator, we see that the energy dispersion is affected by two $\Delta_1$ and $\Delta_2$ spin channels in a fundamentally distinct manner. With the alone singlet or triplet pairing state, what that makes energy spectrum electronically interesting is the fact that the conduction and valence bands touch each other at Dirac point. Whereas, no strikingly say, the band topology of mixed state 3DTI undergoes a change, and a sizeable energy gap is manipulated at Dirac point. This gap can be controlled by tuning the relative magnitude of singlet $|\Delta_s|$ and triplet $|\Delta_p|$ pair potentials. It seems, that the correlations between the spin-singlet and spin-triplet plays the role of effective Dirac mass in the surface states of topological insulator. This can be of an interesting feature of mixed-spin state superconductors in proximity with Dirac-like materials. It is noted that when two spin channels become equal $\Delta_1 = \Delta_2 = 1$, the superconducting excitation reduces to $s$-wave-like one.

However, mass-like gap in Dirac point of surface states can be clearly presented by vanishing quasi-particle wavevector. Consequently, the energy in Dirac point is separated into two parts corresponding two mixed spin channels

$$\mathcal{E}_{mix}(|k|=0) = \begin{pmatrix} \pm\sqrt{\mu_s^2 + |\Delta_1|^2} \\ \pm\sqrt{\mu_s^2 + |\Delta_2|^2} \end{pmatrix}.$$

When we set $|\Delta_1| = |\Delta_s + \Delta_p| = 1$, the size of mass-like gap dependency on the mixed pair potential can be clearly obtained, as shown in Fig. 2. For $\Delta_2 = \pm 1$, we see the energy gap to be closed. This is in agreement with $s$ and $p$-waves superconducting excitations in topological insulators. In the case of $-1 < \Delta_2 < +1$, the gap is immediately opened, and has a maximum when the singlet and triplet state contributions are equal. In the next, we proceed to investigate the dynamical property of such Dirac gap opening via the calculating winding number topological invariant, since it can be of significant importance in topological phase exchange point of view. Moreover, we are interested in zero energy superconductor excitations. Solving Eq. (4) for $\mathcal{E}_{mix.} = 0$, which gives zero subgap energy, results in superconducting gapless state only in the absence of spin-singlet state ($\Delta_s = 0$). To present, in detail, the superconducting zero-energy, we plot in Fig. 3 dispersion energy as a function of $k_x$ and $k_y$ for three cases of singlet, triplet and mixed symmetries of pairing order. When, $\Delta_2 = -1$, which means pure triplet case, the zero-energy occurs. Hence, triplet superconductor components in 3DTI Hamiltonian may give rise to gapless states in Fermi wavevectors $k_F = \sqrt{\mu_s^2 + |\Delta_p|^2}$ (see, Ref. [36]), as shown in Fig. 3(c).

## C  Topological nature of system

In this section, we proceed to investigate the possible topological properties of mixed superconductor state 3DTI in order to achieve an answer for question whether the creation of Dirac mass-like gap is accompanied by topological invariant of band phase exchange. Reaching this goal in our proposed system can be feasible by determining the Berry phase, including the non-trivial topological structure of the wavefunction. Using the wavefunction $|n(\vec{R})\rangle$ of a system, where $\vec{R}$ is the space set of parameters the quantities so-called Berry connection $\mathbf{A}_{\vec{R}}$ and Berry curvature $\mathbf{B}_{\vec{R}}$ are given by

$$\mathbf{A}_{\vec{R}} = -\Im \left\langle n(\vec{R}) \left| \nabla_{\vec{R}} \right| n(\vec{R}) \right\rangle, \quad \mathbf{B}_{\vec{R}} = \nabla_{\vec{R}} \times \mathbf{A}_{\vec{R}}$$

When a system moves along a close path $C$ in the space, the resulting Berry phase $\gamma$ acquired in the wavefunction is

$$\gamma = -\oint_C d\vec{R} \cdot \mathbf{A}_{\vec{R}} = -\int_S d\vec{S} \cdot \mathbf{B}_{\vec{R}}$$

Here, $S$ represents area in the parameter space, enclosed by the contour $C$. In our system, the parameter set is specified by momentum $\vec{k}$, and the Berry phase is also called the Zak phase [38]. Therefore, the topological invariant related to the Zak phase is winding number

$$\omega = \frac{-1}{2\pi} \int_{BZ} dk \, Tr. \left[ \hat{q}^{-1}(k) \, \partial_k \, \hat{q}(k) \right]. \tag{5}$$

The integration is performed over a closed path including wavevectors belonging to the first Brillouin zone. The spectral projection operator $\hat{q}(k)$ defines a map from the reciprocal space in Brillouin zone to the space of unitary matrices belonging to the symmetry group. The $\hat{q}(k)$ matrix is determined via the several constraints concerning to the discrete symmetries imposed on Hamiltonian.

The chiral symmetric Hamiltonian is a needed condition to calculate the winding number, so that it needs to be in block off-diagonal form. Therefore, we may construct formal chiral symmetry operator via the TR and PH symmetry operators (given in the previous section) as following

$$\hat{C} = \hat{\tau}_1 \otimes i\hat{\sigma}_2 = \begin{pmatrix} 0 & 0 & 0 & -i \\ 0 & 0 & i & 0 \\ 0 & -i & 0 & 0 \\ i & 0 & 0 & 0 \end{pmatrix}.$$

Now, we are able to introduce an unitary transformation $\hat{U}_c$, using the eigenvectors of above chiral symmetry matrix

$$\hat{U}_c = \begin{pmatrix} 1 & 0 & 1 & 0 \\ 0 & 1 & 0 & 1 \\ 0 & -i & 0 & i \\ i & 0 & -i & 0 \end{pmatrix},$$

under which the original mixed superconducting DBdG Hamiltonian (2) is transformed to block off-diagonal chiral form

$$\hat{U}_c \mathcal{H}(\mathbf{k}) \hat{U}_c^{-1} = \begin{pmatrix} 0 & 0 & -i\Delta_1 & v_F|k|e^{-i\theta} \\ 0 & 0 & v_F|k|e^{i\theta} & -i\Delta_2 \\ i\Delta_1 & v_F|k|e^{-i\theta} & 0 & 0 \\ v_F|k|e^{i\theta} & i\Delta_2 & 0 & 0 \end{pmatrix}. \tag{6}$$

The energy eigenvalue of filled states is given by

$$\widetilde{E}_{1,2} = -\sqrt{v_F^2|k|^2 + \tilde{\Delta}_1 \pm \sqrt{\tilde{\Delta}_2^2 + v_F^2|k|^2(\Delta_1 - \Delta_2)^2}}.$$

The corresponding chiral eigenstates of Hamiltonian (6) are easily given by

$$|u_1\rangle = \begin{pmatrix} v_F|k|\widetilde{E}_1^{-1}a_1 e^{-i\theta} \\ i\widetilde{E}_1^{-1}b_1 \\ iv_F|k|c_1 e^{-i\theta} \\ 1 \end{pmatrix}, \quad |u_2\rangle = \begin{pmatrix} i\widetilde{E}_1^{-1}b_2 \\ v_F|k|\widetilde{E}_1^{-1}a_2 e^{i\theta} \\ 1 \\ iv_F|k|c_1 e^{i\theta} \end{pmatrix},$$

$$|u_3\rangle = \begin{pmatrix} v_F|k|\widetilde{E}_2^{-1}a'_1 e^{-i\theta} \\ i\widetilde{E}_2^{-1}b'_1 \\ iv_F|k|c_2 e^{-i\theta} \\ 1 \end{pmatrix}, \quad |u_4\rangle = \begin{pmatrix} i\widetilde{E}_2^{-1}b'_2 \\ v_F|k|\widetilde{E}_2^{-1}a'_2 e^{i\theta} \\ 1 \\ iv_F|k|c_2 e^{i\theta} \end{pmatrix}, \tag{7}$$

where

$$a_{1(2)} = 1 + \Delta_{1(2)}c_1, \quad b_{1(2)} = -\Delta_{2(1)} + v_F^2|k|^2 c_1,$$
$$a'_{1(2)} = 1 + \Delta_{1(2)}c_2, \quad b'_{1(2)} = -\Delta_{2(1)} + v_F^2|k|^2 c_2,$$
$$c_{1(2)} = \frac{\Delta_2 - \Delta_1}{A_{1(2)}}, \quad A_{1(2)} = \Delta_1^2 + v_F^2|k|^2 - \widetilde{E}_{1(2)}^2.$$

To facilitate the calculation, it helps to introduce the projection operator $\hat{p}(k) = \sum_{i \in filled} |u_i\rangle \langle u_i|$. For what follows, it is convenient to introduce the $Q$ matrix by $\hat{Q}(k) = 2\hat{p}(k) - \hat{1}$. Corresponding to the block-off-diagonal chiral symmetric Hamiltonian (6), the $Q(k)$ matrix is also block-off-diagonal:

$$\hat{Q}(k) = \begin{pmatrix} 0 & \hat{q}(k) \\ \hat{q}^\dagger(k) & 0 \end{pmatrix}.$$

Here 0 is a zero matrix with 2 by 2 dimension and $q(k)$ is the off-diagonal component of $Q$ matrix. There can be a topological invariant, which is obtained only in the presence of a symmetry. Indeed, the chiral symmetry gives rise to result in winding number topological invariant. We are now set to calculate the topological invariant winding number via the $\hat{q}(k)$ matrix. Having more complicated chiral eigenfunctions (7), we try to find a huge expression for $\hat{q}(k)$ matrix, and neglect to write it here. Inevitably, the numerical method may be used to find the winding number. The analytical expression for the off-diagonal block of spectral projector matrix is unwieldy, and further treatment about evaluating the topological invariant of this system are processing now.

### D  Effective subgap

To more clarify the mixed superconducting state exhibition, we focus on superconductor effective gap originated from singlet and triplet correlations. Actually, magnitude of mixed effective gap depends on the relative amplitude between the singlet and triplet components, which can control the height of forming subgap at the interface, playing a crucial role in hole-reflection for incident electrons. In order to derive the exact form of effective gap, we need to refer energy spectra of topological insulator in proximity of a $s$-wave and $p$-wave superconductor, separately [36]

$$\epsilon_{(s-wave)}^2 - |\Delta_s|^2 = (v_F|k| \mp \mu_s)^2,$$

$$\epsilon_{(p-wave)} = \pm v_F|k| \mp \sqrt{\mu_s^2 + |\Delta_p|^2}.$$

We reconstruct the energy spectra of Eq. (4) as

$$\mathcal{E}_{mix.}^2 - |\Delta_{eff}|^2 = \left(v_F|k| \pm \frac{\sqrt{\varepsilon_R}}{2v_F|k|}\right)^2,$$

in order to exploit an exact expression for effective mixed gap as following:

$$|\Delta_{eff}| = \sqrt{\tilde{\Delta}_1 - \frac{|\Delta_1 - \Delta_2|^2}{4} - \frac{\tilde{\Delta}_2^2}{\mu_s' + \sqrt{\mu_s'^2 + 4\tilde{\Delta}_2^2}}}. \tag{8}$$

The normalized chemical potential $\mu_s' = 2\mu_s^2 + \frac{1}{2}|\Delta_1 - \Delta_2|^2$ indicates mixed spin channels.

The position of superconducting gap in $\Gamma$ point corresponds to the relation $\frac{1}{2}(\mu_s' + \sqrt{\mu_s'^2 + 4\tilde{\Delta}_2^2})^{1/2}$. It should be noted that in the limit of alone singlet or triplet case, the $\Gamma$ point only depends on $\mu$, while in the case of mixing potential, existence of two mixed components $\Delta_1$ and $\Delta_2$ causes to shift the position of superconducting gap. We find $\Delta_{eff}$ to become zero in the absence of spin-singlet contribution achieving by $\Delta_1 = -\Delta_2$. Interestingly, in the lake of spin-triplet contribution, which is obtained by $\Delta_1 = \Delta_2$, the effective gap is clearly reduced to the isotropic order parameter. This is completely in agreement with previously reported results, that the former corresponds to the gapless topological insulator superconductor state, and the latter means conventional $s$-wave superconducting excitations one. The behavior of these cases is shown in Fig. 4.

## III  TRANSPORT PROPERTIES

### A  Andreev tunneling conductance

In this section, we will focus on the transport properties of the simplest hybrid normal/superconductor structure deposited on top of a topological insulator in order to investigate how Andreev reflection and conductance spectroscopy are influenced by the superconducting mixed order parameter. The unconventional mixed superconductivity in TIs should manifest itself in the observable phenomena at the boundaries of a hybrid structure. We analyze Andreev reflection probability in the surface states by employing a scattering matrix formulation along the lines of Blonder-Tinkham-Klapwijk (BTK) theory. To this end, let us now proceed to introduce the eigenstates of Hamiltonian (2). The wave function in the topological insulator mixed superconducting is achieved from a set of $4 \times 4$ coupled matrix equations. Here, there are four unknowns to derive the eigenfunction in the electron-hole basis (Nambu basis), $\psi_{mix.} = \left[\psi_{k\uparrow}, \psi_{k\downarrow}, \psi_{-k\uparrow}^\dagger, \psi_{-k\downarrow}^\dagger\right]$. The normalization condition, $|\psi_{k\uparrow}|^2 + |\psi_{k\downarrow}|^2 + |\psi_{-k\uparrow}^*|^2 + |\psi_{-k\downarrow}^*|^2 = 2$ is used to conserve the intensity of the edge states. From equation $\mathcal{H}\psi_{mix.} = \mathcal{E}_{mix.}\psi_{mix.}$, we, after cumbersome analytical calculations, express eigenfunction of a electron(hole)-like quasiparticle states in terms of following equation:

$$\psi_{mix.} = \sqrt{\frac{2}{\mathcal{A}}} \begin{pmatrix} \mathcal{M}_1 \\ v_F|k|\mathcal{M}_2 e^{i\theta} \\ v_F|k|\mathcal{M}_3 e^{i\theta} \\ \mathcal{M}_4 \end{pmatrix}, \tag{9}$$

where $\mathcal{A}$ is the normalization constant and

$$\mathcal{M}_1 = \Delta_1 |\Delta_2|^2 + \Delta_2 v_F^2 |k|^2 + \Delta_1(\mu_s^2 - \mathcal{E}_{mix.}^2),$$

$$\mathcal{M}_2 = \Delta_1(\mu_s - \mathcal{E}_{mix.}) + \Delta_2(\mu_s + \mathcal{E}_{mix.}),$$

$$\mathcal{M}_3 = \Delta_1 \Delta_2^* + v_F^2 |k|^2 - (\mu_s + \mathcal{E}_{mix.})^2,$$

$$\mathcal{M}_4 = (\mu_s + \mathcal{E}_{mix.})\left(|\Delta_2|^2 + \mu_s^2 - \mathcal{E}_{mix.}^2 + v_F^2|k|^2\right) - 2\mu_s v_F^2 |k|^2.$$

Due to relativistic dynamics, two independent spin channels $\Delta_1$ and $\Delta_2$ are simultaneously appeared in the wave function. Because the motion of quasiparticles is determined by incidence angle $\theta$, the resulting wave function is related to the direction of motion. If we define angle $\theta$ for right movers, then left movers is described by $\pi - \theta$. Accordingly, pairing potentials spatially depend on direction of motion. Therefore, two spin channels in Eq. (9) are defined only for right movers, and we can replace them for left movers by $\Delta_{1(2)} \to \Delta_{2(1)}$ (see Fig. 1(a)). Also, the explicit wavevector of quasiparticles in terms of superconducting excitation energy and mixed-spin channels is given by

$$|k| = \mathcal{E}_{mix.}^2 + \mu_s^2 - \Delta_1 \Delta_2 - \sqrt{4\mathcal{E}_{mix.}^2 \mu_s^2 + \mathcal{E}_{mix.}^2 |\Delta_1 - \Delta_2|^2 - \mu_s^2 |\Delta_1 + \Delta_2|^2}.$$

To accommodate superconductivity by means of the proximity effect experimentally, it is necessary to realize the condition $\mu_s \gg |\Delta_{1,2}|$ to have a sufficiently large density of states. In this way, a superconductor electrode deposited on top of the topological insulator would be suitable experimentally, as Fig. 1(b).

The total wave function in the normal region of junction ($x < 0$) by regarding two possible fates upon scattering, normal and Andreev reflections of an incident electron, may then be written as:

$$\Psi_N = e^{ik_N^y y} \left( e^{ik_N^e x} \begin{bmatrix} 1 & e^{i\alpha} & 0 & 0 \end{bmatrix}^T + \right.$$

$$\left. re^{-ik_N^e x} \begin{bmatrix} 1 & -e^{-i\alpha} & 0 & 0 \end{bmatrix}^T + r_A e^{ik_N^h x} \begin{bmatrix} 0 & 0 & 1 & -e^{-i\alpha_h} \end{bmatrix}^T \right), \quad (10)$$

where $\alpha$ and $\alpha_h$ denote the electron and hole angles of incidence, while $r$ and $r_A$ are the normal and Andreev scattering coefficients, respectively. Due to the broken translational symmetry, the $x$-component of the momentum in normal region ($k_N^x$) is non-conserved, whereas $y$-component ($k_N^y$) is conserved, and can be acquired from normal region eigenstate. The Fermi momentum in the normal and superconducting part of the system can be controlled by means of chemical potential in each region. Setting up the scattering wavefunctions and utilizing appropriate boundary condition, $\Psi_N = \Psi_S$ at $x = 0$, where $\Psi_S = t^e \psi_{mix.}^e + t^h \psi_{mix.}^h$, one is able to extract the normal and Andreev reflection coefficients, which depend on the angle of incidence and the mixed state channels excitation energy. We find following solutions for normal and Andreev reflection coefficients:

$$r = \Gamma \left[ \mathcal{M}_1^e \eta_4 - \mathcal{M}_1^h \eta_3 \right] - 1, \tag{11-a}$$

$$r_A = \Gamma \left[ \mathcal{M}_3^e \eta_4 - \mathcal{M}_3^h \eta_3 \right], \tag{11-b}$$

where we have introduced

$$\Gamma = \frac{2 \cos \alpha}{\eta_1 \eta_4 - \eta_2 \eta_3},$$

$$\eta_{1(3)} = \mathcal{M}_{2(4)}^e + \mathcal{M}_{1(3)}^e e^{-i\alpha(\alpha_h)},$$

$$\eta_{2(4)} = \mathcal{M}_{2(4)}^h + \mathcal{M}_{1(3)}^h e^{-i\alpha(\alpha_h)}.$$

It follows, according to the BTK formalism [39], the normalized conductance ($G/G_0$) can be calculated,

$$G/G_0 = \int_{-\pi/2}^{\pi/2} d\alpha \, \cos \alpha \left[ 1 + |r_A|^2 - |r|^2 \right], \tag{12}$$

and the normalization constant is chosen as $G_0 \approx N(E_F)we^2/\pi\hbar^2$, where $N(E_F) \approx E_F/2\pi(\hbar v_F^2)$ is density of state with $w$ being width of the junction.

As we show now, the effect of the two distinct spin channels can be nicely seen in the experimentally accessible electrical conductance. In Fig. 5(a), we plot the subgap conductance spectra of the NS structure resulting from the Andreev process, calculated with superconductor and normal region chemical potentials $\mu_s = 10\Delta_m$ and $\mu_N = 1\Delta_m$, respectively. The maximum suppression of conductance happens for the case of opposite spin channels, $\Delta_1 = -\Delta_2$. In this case, however, it seems that the appearance of unconventional superconductivity is manifested by an enhancement of the zero-bias conductance peak.

Importantly, in the mixed state range $0 < \Delta_2 < 1$, the two coherence conductance peaks exist at $\mathcal{E}_{mix.} = \Delta_{eff}$, and a transition of conductance peak into the zero-bias conductance can also be achieved by increasing the amplitude of $\Delta_2$. By focusing on the effective gap relation, Eq. (8), one can find $\Delta_{eff} \cong \Delta_s$ for high values of chemical potential of superconductor. Therefore, in order to preserve mixed state subgap effect, we may apply minimum possible doping, where the condition $|\Delta_m| \ll \mu_s, E_F$ is still satisfied. Stehno et al have reported the experimental implementation of this scenario [40]. The presence of $s$-wave pairing with subdominant $p$-wave admixture order parameter has been predicted on $Nb$/topological insulator/$Au$ devices, where the topological insulator is either alloyed $Bi_{1.5}Sb_{0.5}Te_{1.7}Se_{1.3}$ or $BiSbTeSe_2$. Indeed, the conductance dips at the induced-gap value and the increased conductance near zero energy in above both spectra of samples, can be explained by the dominant triplet superconducting components in 3DTI [40]. In analogy, in NCS superconductors with broken inversion symmetry, the transport signatures in N/S junction depend on the degree of mixing of singlet and triplet pair potentials. In Ref. [25], Burset et al have analyzed tunneling conductance of normal/noncentrosymmetric superconductor junction, and reported a zero-bias conductance peak for the case $\Delta_s < \Delta_p$, analogous to our finding, here.

In Fig. 5(b), we present the signature of doping level of N region in resulting normal conductance and formation of zero-bias conductance peak. A sharp conductance peak in zero-bias can be nicely seen in a low doping, whereas the zero dip of conductance is appeared by increasing the normal region doping. It is interesting to note that the conductance peaks can also be controlled by changing the pairing potential admixture. For comparison, we have included in Fig. 5(c) the conductance of junction with two possible $p$-wave symmetry functions. For $\mathbf{d(k)} = \Delta_p \cos\theta\hat{z}$, the conductance peaks located at the effective mixed gap is smaller than that for $\mathbf{d(k)} = \Delta_p \sin\theta\hat{z}$. This scenario becomes completely vice versa for resulting zero-bias conductance. To more clarify the signature of two mixed-state channels in conductance peak displacement, we present, in Fig. 5(d), subgap conductance curve in terms of $\Delta_2$ and bias energy. This figure clearly demonstrates conductance peak displacement towards zero-bias with the increase of magnitude of triplet pair potential.

## B  Andreev Bound States in Josephson junction

We now consider the strictly one-dimensional superconductor/insulator/superconductor (S/I/S) Josephson junction in the $x$-direction on the surface of 3D topological insulator, as sketched in Fig. 1(c). The measurement of the supercurrent which is carried by Cooper pairs can be one of the useful tools to reveal effective symmetry manipulated by inducing an actual superconductivity. The mixed superconductivity in topological insulator particularly manifests itself in the Josephson effect. The pairing potential vanishes in the insulator middle region and is nonzero in the two superconductor terminals. The order parameter is assumed to have different phases and the same amplitude in the left and right superconductors. The insulator region length $L$ (distance between two superconductor terminals) is assumed to be much smaller than the superconducting coherence length $\xi = \hbar v_F/\Delta$. For make contact with experimental parameters, the junction length should be smaller than $0.7~\mu m$. We introduce a gate potential $U_0$ for insulator region for the possibility of electron scattering in the junction. In this condition, the main interesting Klein tunneling effect takes place between the terminals that is independent of the barrier shape.

The wave function in the insulator region remains the same to normal region while we re-label $\Psi_S \to \Psi_S^r$ and concomitantly $\{t^e, t^h\} \to \{t^{er}, t^{hr}\}$ for the right superconductor region $x > L$. The pairing potential is a combination of singlet and triplet states which adopts the following form for each left and right S regions

$$\Delta_{1,2} = \begin{cases} \left(\Delta_s \pm \Delta_p\, e^{i\theta}\right) e^{i\varphi_r}, & x > L \\ \left(\Delta_s \mp \Delta_p\, e^{i\theta}\right) e^{i\varphi_l}, & x < 0 \end{cases}. \qquad (13)$$

The pairing potential is assumed to have different phases in the left and right regions, and the current flowing the Josephson junction depends on the phase difference $\Delta\varphi = \varphi_r - \varphi_l$. It, then, remains to introduce the wave function for the left superconductor region ($x < 0$), which reads $\Psi_S^l = t^{el}\psi_{mix.}^e + t^{hl}\psi_{mix.}^h$. To identify the energy spectrum for the Andreev bound state, we match the wave functions around $x = 0$, which yields

$$\Psi_{S_{x\to 0}}^l = \begin{pmatrix} 1 + iZ\sigma_1 & 0 \\ 0 & 1 - iZ\sigma_1 \end{pmatrix} \Psi_{S_{x\to 0}}^r, \qquad (14)$$

where the barrier strength is defined as dimensionless parameter $Z$. By inserting the superconducting wave functions into Eq. (14), we arrive at four linear algebraic equations for the four constants $t^{er}$, $t^{hr}$, $t^{el}$ and $t^{hl}$. For the case of $\Delta_p = 0$ and $\Delta_s = 1$, which we have no longer mixed state, the ABS solutions arrive at the well known previously reported equation [16]. When the spin state is mixed, finding the analytical expression for ABS becomes impossible. The cumbersome and time-consuming analytical calculations has been done in this relation, and finally, from Eq. (14), we obtain an equation

$$e^{-2i\varphi}\mathcal{G}_1 + \mathcal{G}_2 = 0,$$

where $\mathcal{G}_1$ and $\mathcal{G}_2$ are more complicated functions of bound energy, barrier parameter and incidence angle. We can numerically obtain ABS spectrum as a function of superconducting phase difference $\Delta\varphi$ and propagation angle $\theta$.

We show that the same outcomes similar to those previously obtained for Josephson effect in topological insulator with alone $s$-wave symmetry are achieved [41]. The $4\pi$-periodic gapless bound energies in normal incidence $\theta = 0$ are appeared, which are protected by the TR symmetry (see, Fig. 6(a)). Also, these states correspond to the chiral Majorana bound energy modes, so that the energy curves of electron and hole are continuously connected. The range of superconductor state admixture is controlled by the magnitude of spin channel $\Delta_2$, where we take the other mixed spin channel to be unit, $\Delta_1 = 1$. Hence, when $\Delta_2$ is varied from 1 to $-1$, the mixed state level is continuously changed from $s$-wave symmetry to $p$-wave one. Independent of admixture level tuned by $\Delta_2$, the ABS spectra exhibits zero energy and maximum slope for superconductor phase difference $\Delta\varphi = (2n + 1)\pi$ ($n$ is integer number). Whereas, $\Delta\varphi = 2n\pi$ results in flat energy curve. It is noticed, that the amplitude of ABS oscillations significantly diminishes in the mixed spin state. These features are presented in Fig. 6(a), where ABS plot are given as a function of phase difference for the superconductor chemical potential and middle region insulator strength parameter magnitudes $\mu_s = 15$ and $Z = 0.5$, respectively.

For the critical case of mixed spin channel $\Delta_2 = -1$, which our Josephson junction will be in pure spin-triplet symmetric state, the ABS curvature goes to flattening. These behaviors of mixed superconducting ABS can be originated from Dirac band gap creation and strongly effective subgap decreasing in the system. Furthermore, in Fig. 6(b), we plot bound state energy for finite angle of incidence as a function of phase difference. As expected, the signature of nonzero incidences of quasiparticles to the superconductor/insulator interface is observed as vanishing chiral Majorana mode via the opening a large gap in ABS. Consequently, the period of ABS oscillations becomes $2\pi$ in the presence of a momentum mismatch, which is due to finite backscattering. The decrease of the amplitude of ABS is determined by the mixing level and the angle of incidence. It should, however, be noted, that the change of amplitude of ABS curves with the incidence angle strongly depends on the magnitude of $\Delta_2$. We show increasing the angle of incidence in the range from 0 to $0.2\pi$ enhances the value of the bulk gap from 0 to 0.3 for the mixing state ($\Delta_2 = 0.2$), whereas for $\Delta_2 = 1$ ($s$-wave superconductivity dominant case), it takes place from 0 to 0.5.

## C   $0 - \pi$ supercurrent

After the ABS spectrum is found, we numerically calculate the angle-averaged supercurrent that is given by the phase difference dispersion of bound state energy $E(\Delta\varphi)$. The normalized Josephson current in the short junction case can be calculated according to the standard expression [42],

$$I/I_0 = \int_{-\pi/2}^{\pi/2} d\theta \cos\theta \, \tanh\left(\frac{E(\Delta\varphi)}{2K_B T}\right) \frac{dE(\Delta\varphi)}{d\Delta\varphi} \quad (15)$$

where $I_0 = \frac{e|k|W\Delta_m}{\pi\hbar}$ is the normal current in a sheet of TI of width $W$, $K_B$ and $T$ are the Boltzmann constant and temperature, respectively. In Fig. 7(a), the Josephson current as a function of superconducting phase difference is demonstrated for several magnitudes of $\Delta_2$. As a usual result in similar systems, the $2\pi$-periodic current-phase curve is found for every admixture level, in spite of the presence of the spin-triplet component of the pair potential. The main difference between the mixed-spin channel and pure spin-singlet one ($\Delta_2 = 1$), as shown in Fig. 7(a), is that the Josephson supercurrent is strongly suppressed as the amplitude of the spin-triplet contribution grows upto $\Delta_p = 0.8$. In Fig. 7(b), we repeat the previous calculation of Josephson current for different values of insulator barrier strengths $Z$. Here, it occurs an interesting scenario, where the exact sinusoidal curve of supercurrent is achieved in the case of large $Z$. According previous specific work [43], the abrupt crossover phase-current curve originated from the gapless ABS is observed in the low value of barrier strength. Finally, to clarify the signature of mixed superconductivity on the critical current, which is defined as the maximum of Josephson current, we analyze numerically and plot the barrier strength dependence of critical current. Figure 7(c) shows the normalized critical current $I_c/I_0$ for different magnitudes of mixed-spin characteristic parameter $\Delta_2$. We show that the critical current strongly decreases with the increase of spin-triplet contribution. The reason of this effect may be described by decreasing the effective superconducting gap in the mixed state.

## IV   SUMMARY AND CONCLUSIONS

In summary, from a more fundamental perspective, the distinction between the energy spectrum in the mixed-spin state superconductors and surface states of topological insulators teaches us something new about the interplay between mixed state of superconductivity and topologically protected by time-reversal symmetry Dirac-like fermions. In one hand, the inversion symmetry breaking in a noncentrosymmetric superconductor, and gapless surface state resulted from spin-orbit coupling on the other hand, can be strongly inter-correlated to capture the new effects in spin magnetization and spin transportation. Magnetoelectric effect caused by supercurrent in NCS superconductors has been reported, recently [44]. There is a delicate point, that the Hamiltonian of two-dimensional NCS superconductors is decoupled into two spin channels $\Delta_1$ and $\Delta_2$ with different energies. Whereas, in the presence of topological insulator, two spin channels are strongly coupled with the same energy, and both right-moving and left-moving electron(hole) quasiparticles may experience the two spin channels.

In this paper, we have analyzed the effect of proximity-induced mixed spin-singlet and spin-triplet symmetry on the surface states of a topological insulator. The particle-hole and chiral symmetric properties of Dirac-Bogoliubov-de Gennes Hamiltonian has been investigated to capture the topology class. We have introduced the new spin channels $\Delta_1$ and $\Delta_2$ for mixed state in the presence of topological insulator. Particularly, we have found a transformation matrix, under which the Hamiltonian is diagonalized, and, interestingly, the new mixed-spin channels were located at the diagonal elements. Consequently, it is formally expected to appear a Dirac mass-like gap in the surface states. This can be considered as a key feature of the present structure. It is noticed that there exist similar situation, when a magnetization in $z$-direction $m_z$ is induced to 3DTI [16]. Therefore, one can report that mixed-state superconductivity induction may play simultaneous role of magnetic field appearance in topological insulators. This is considered a significant feature, particularly in NCS superconductors [23, 45, 46, 47]. Next, we have further tried to clarify possible phase transition from original gapless in conventional superconducting to gaped surface states in unconventional mixed one via the evaluating the topological invariant winding number for the chiral eigenstates. To this end, we have constructed a chiral unitary matrix, under which the Hamiltonian is transformed to its block-off-diagonal form. Because the spectral projection matrix $\hat{q}(k)$ has been obtained in an unwieldy analytical expression, then the winding number will be presented in another work.

Regarding the fact that superconducting electron-hole excitation in topological insulator is gapless with $p$-wave pairing symmetry, it was necessary to reveal the effective subgap in the mixed state case, which is identified to have a complicated dependency on mixed spin channels. However, we see a sizable subgap on the Fermi surface, and it diminishes when the $p$-wave symmetry contribution is dominated. We have thus systematically proceeded to investigate the characteristic transport properties for subgap tunneling in N/S and Josephson S/I/S junctions. Our proposal has clear advantages in experimental accessibility. The Josephson current on the surface of the 3DTI has been experimentally observed, where Josephson junction $Nb$/epitaxial $(Bi_{0.5}Sb_{0.5})_2Te_3/Nb$ in the two steps have been fabricated and good $I-V$ characteristics presented [11]. Also, tunneling conductance spectroscopy has been performed across hetero-$Nb$/topological insulator/$Au$, recently [40].

**Acknowledgements**

The authors would like to thank Vice-presidency and also National Elites foundation of I.R. of Iran under grant number 15/295 for supporting the present work and post-doctorate course of MK at the Urmia University.

**Figure captions**

**Figure 1(a), (b), (c), (d)** (color online) (a) A schematic of two coupled mixed-spin channels $\Delta_1$ and $\Delta_2$ for right-left-moving quasiparticles, (b) and (c) sketch of the topological insulator-based N/S subgap tunneling and S/I/S Josephson junctions with mixed-spin state superconductivity, (d) A polar plot of two spin channels $\Delta_1$ and $\Delta_2$ with the incidence angle is shown for three different values of $\Delta_s$ and $\Delta_p$. The solid lines denote the $\Delta_1$, while the crossed lines correspond to the $\Delta_2$. Black curves represent pair potential for $|\Delta_s| = 0.8$, $|\Delta_p| = 0.2$, violet curves for $|\Delta_s| = |\Delta_p| = 0.5$ and blue curves for $|\Delta_s| = 0.2$, $|\Delta_p| = 0.8$.

**Figure 2** (color online) The excitation spectra in superconductor topological insulator, calculated from Eq. (4). The mixed superconducting state features a mass-like gap in topological insulator.

**Figure 3(a), (b), (c)** (color online) Contour plot of superconducting excitation spectra on the surface state of topological insulator for (a) $|\Delta_2| = 0$, (b) $|\Delta_2| = 1$ and (c) $|\Delta_2| = -1$. It is seen that the superconducting zero energy only occurs in $|\Delta_2| = -1$.

**Figure 4** (color online) Effective band gap resulting from Eq. (8) with $\mu_s = 0.1\Delta_m$ as a function of $\Delta_1$ and $\Delta_2$.

**Figure 5(a), (b), (c), (d)** (color online) Normalized tunneling conductance versus bias voltage in N/S junction. The plots in (a) show the results for different values of $\Delta_2$ and the effect of the pairing potential is indicated, (b) show the results for different values of $\mu_N$ and (c) indicate results for two different symmetry of $p$-wave functions. The tunneling conductance versus bias voltage and spin channel $\Delta_2$ is plotted in (d).

**Figure 6(a), (b)** (color online) (a) Plot of the Andreev bound state energy versus phase difference when $\theta = 0$. The black solid line corresponds to $|\Delta_2| = 1$, blue solid line to $|\Delta_2| = 0.6$, red solid line to $|\Delta_2| = 0.2$ while the green dashed line corresponds to $|\Delta_2| = 0$. Also, the pink dashed line corresponds to $|\Delta_2| = -0.2$, light blue dashed line to $|\Delta_2| = -0.6$ and light green dashed line to $|\Delta_2| = -1$, (b) Plot of the Andreev bound state energy versus phase difference for various values of $\theta$. The black curves correspond to $|\Delta_2| = 1$ and pink curves to $|\Delta_2| = 0.2$. Dependence of the Andreev bound state energy for diferent values $\theta = 0$ (pink solid line), $\theta = 0.1\pi$ (pink dashed line), $\theta = 0.2\pi$ (pink dash-dotted line) and $\theta = 0.25\pi$ (pink dotted line) when $\mu_s = 15\Delta_m$ and $Z = 0.5$ is shown.

**Figure 7(a), (b), (c)** (color online) Plot of the normalized Josephson supercurrent as a function of the phase difference with respect to varying (a) $\Delta_2$ and (b) barrier parameter $Z$ when $\mu_s = 10\Delta_m$. We have set $Z = 0.25$ for (a) and $|\Delta_2| = 0.6$ for (b). Plot (c) represents the critical current versus barrier strength when $\mu_s = 10\Delta_m$.

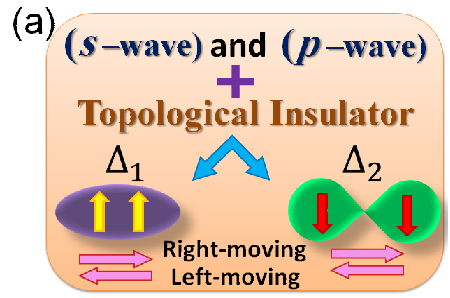
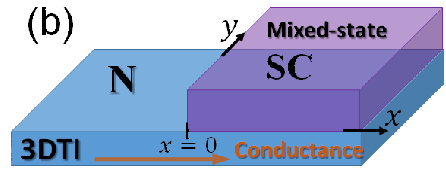
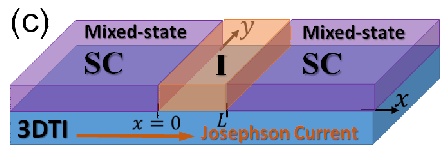
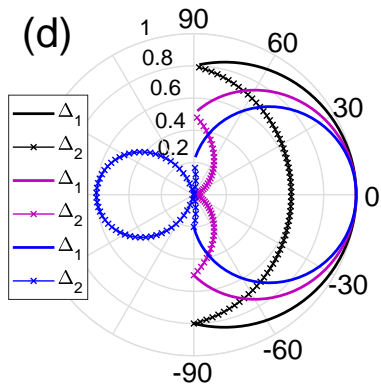

Figure 1: (a), (b), (c), (d)

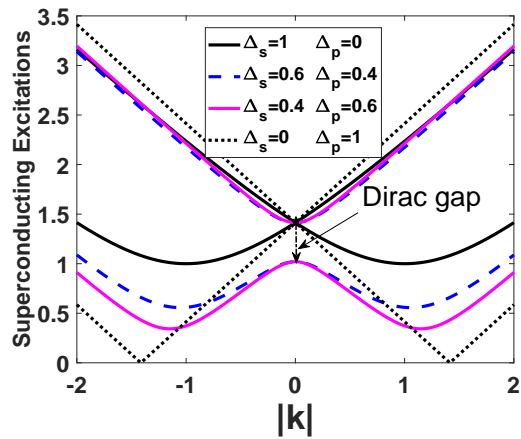

Figure 2:

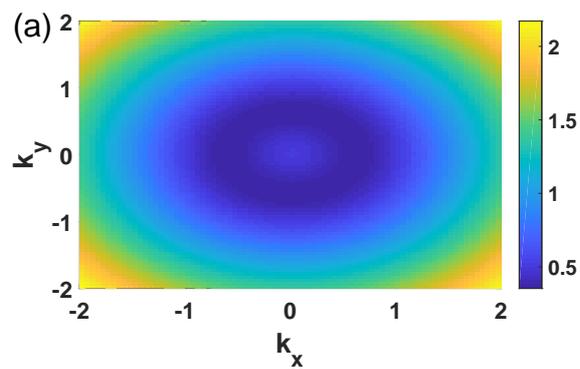

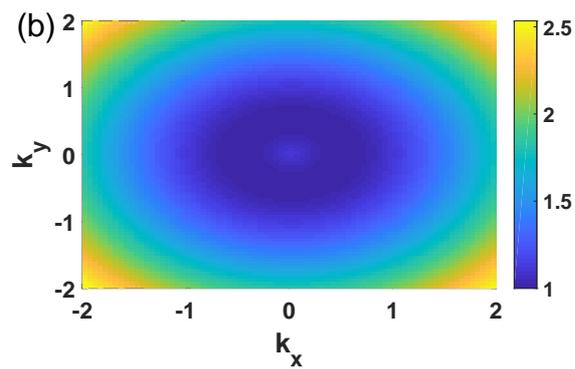

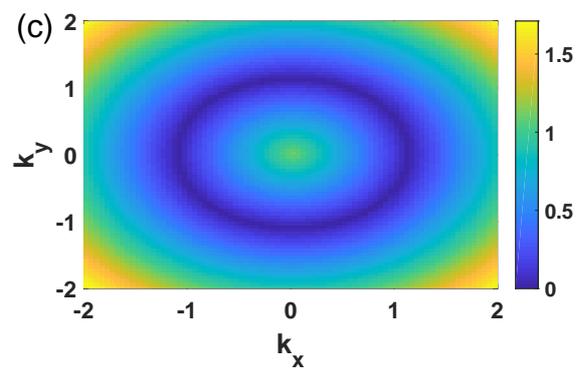

Figure 3: (a), (b), (c)

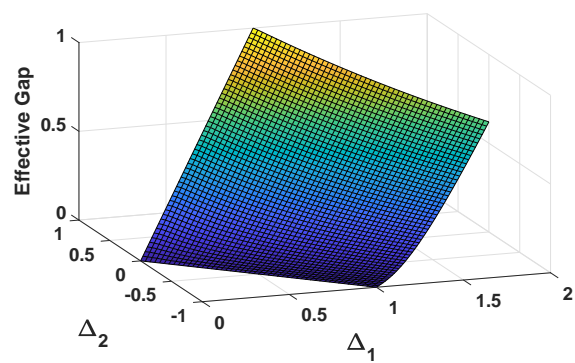

Figure 4:

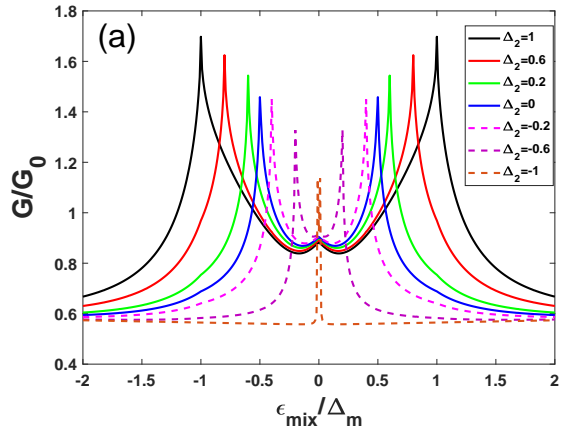

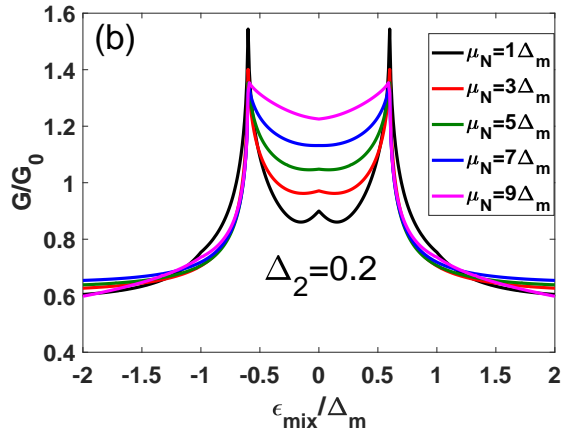

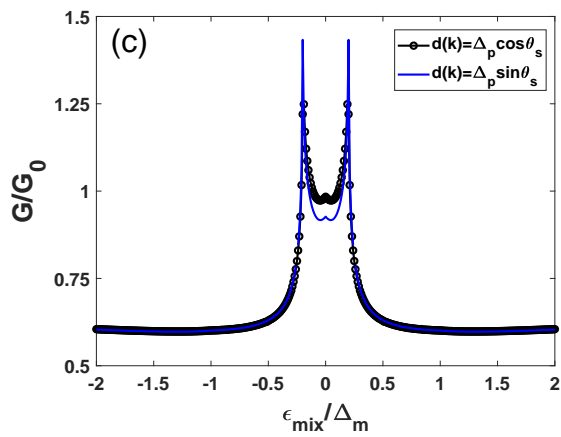

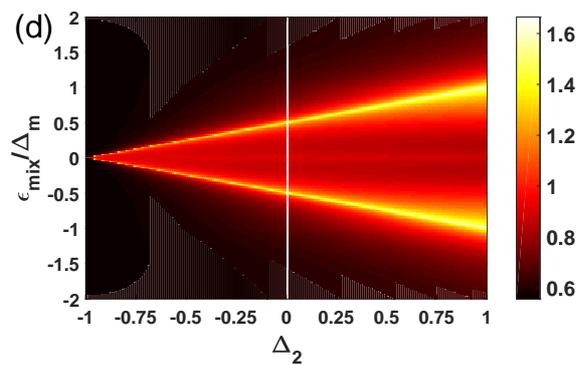

Figure 5: (a), (b), (c), (d)

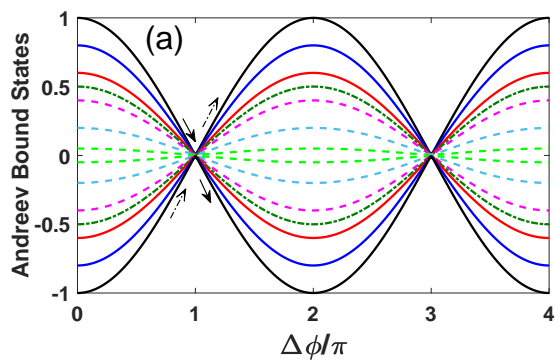

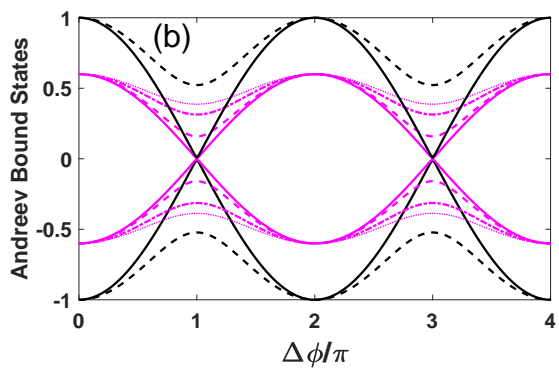

Figure 6: (a), (b)

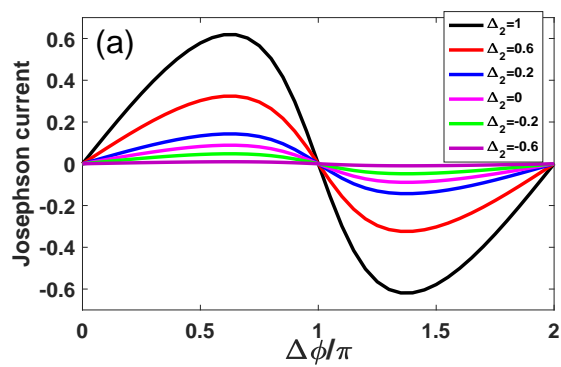

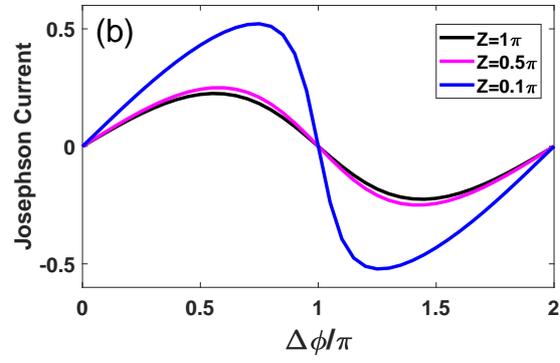

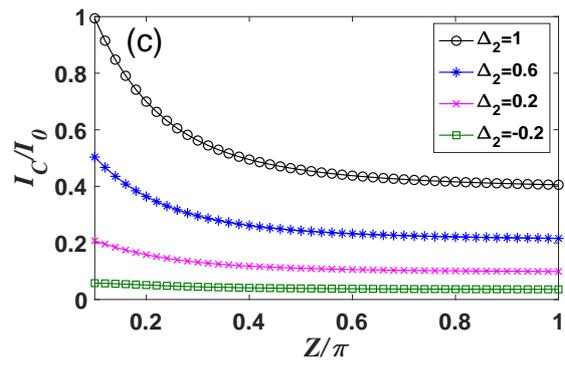

Figure 7: (a), (b), (c)